\documentclass[aps,pra,superscriptaddress,floatfix,twocolumn,a4paper,showpacs,tightenlines]{revtex4}

\usepackage{graphicx}
\usepackage{amsmath}
\usepackage{amssymb}
\usepackage{epsf,latexsym}
\usepackage{times}
\usepackage{xspace}
\newcommand{\IPEX}[3] {\ensuremath{\left\langle {#1} {\left| {#2} \right|} {#3}\right\rangle}}
\newcommand{\PAPER}{paper\xspace}
\newcommand{\OP}[2] {\ensuremath{\left| {#1} \right\rangle \left\langle{#2}\right|}}
\newcommand{\half} {\ensuremath{\frac{1}{2}}}

\newcommand{\ket}[1]{\ensuremath{\left| #1 \right\rangle}}
\newcommand{\bra}[1]{\ensuremath{\left\langle #1 \right|}}

\newcommand{\EX}[1] {\ensuremath{\left\langle #1 \right\rangle}}
\newcommand{\be}{\begin{equation}}
\newcommand{\bel}[1]{\begin{equation}\label{#1}}
\newcommand{\ee}{\end{equation}}
\newcommand{\ba}{\begin{eqnarray}}
\newcommand{\ea}{\end{eqnarray}}

\newcommand{\Fig}[1]{Fig.~\ref{#1}}
\newcommand{\Eq}[1]{Eq.~(\ref{#1})}

\newcommand{\LEEDS}{Quantum Information Science, School of Physics and Astronomy, %
University of Leeds, Leeds LS2 9JT, UK.}
\newcommand{\LBRO}{Department of Physics, Loughborough University,
                Loughborough, Leics LE11 3TU, United Kingdom}
\newcommand{\NTT}{NTT Basic Research Laboratories, NTT 
Corporation, 3-1 Morinosato-Wakamiya, Atsugi, Kanagawa 243-0198, Japan.}
\begin{document}


\title{Overcoming decoherence in the collapse and revival of spin Schr\"{o}dinger-cat states}

\author{M. J. Everitt}
\email{m.j.everitt@physics.org}
\affiliation{\LBRO}
\author{W. J. Munro}
\affiliation{\NTT}
\affiliation{\LEEDS}
\author{T. P. Spiller}
\affiliation{\LEEDS}

\date{\today}

\begin{abstract}
In addition to being a very interesting quantum phenomenon, Schr\"{o}dinger-cat state swapping has the potential for application in the preparation of quantum states that could be used in metrology and other quantum processing. We study in detail the effects of field decoherence on a Schr\"{o}dinger-cat state-swapping system comprising a set of identical qubits, or spins, all coupled to a field mode. We demonstrate that increasing the number of spins actually mitigates the effects of field decoherence on the collapse and revival of a spin Schr\"{o}dinger-cat state, which could be of significant utility in quantum metrology and other quantum processing.
\end{abstract}



\maketitle

Two of the most peculiar and distinctive phenomena of quantum mechanics are encapsulated in entanglement and macroscopically distinct superpositions of states (or Schr\"{o}dinger-cat states) -- both of which can be used to realise in Einstein's words a \emph{``spooky action at a distance''}. With applications ranging from metrology, information processing, communication to computation these phenomena account for the most powerful and interesting aspects of quantum mechanics~\cite{wheeler83,plk,Pen98,NC2000,Munro2002, Ralph2003}. Given current experimental progress, at present there is significant interest in quantum technologies that can offer advantage with modest quantum resources, such as metrology. Here, sensing beyond the standard quantum limit can be achieved using entangled resources, or Schr\"{o}dinger-cat states. One candidate tool for the preparation of desired resources could be the use of Schr\"{o}dinger-cat state swapping, where a Schr\"{o}dinger-cat state is transferred between a field and a system of spins. For example, a field Schr\"{o}dinger-cat state might be prepared through interaction with a single qubit or spin, or by some other means, and then swapped into a multi-spin system for use in metrology. 

Clearly, a practical concern with such operations is the effect of decoherence. Here we examine the effects of field decoherence on a Schr\"{o}dinger-cat state swapping system of $N$ spins all coupled individually to a quantum field mode. Such a set-up might be realised by systems as diverse as atoms in a cavity through to a set of superconducting qubits coupled to a strip line resonator. For metrological and other applications, larger $N$ resources offer improved quantum advantage. It is often the case that increasing the number of qubits results in higher susceptibility to decoherence. However, here we show the reverse holds in this Schr\"{o}dinger-cat state-swapping scenario and that pursuing the desirable goal of increasing the number of qubits actually mitigates against decoherence.

We begin our discussion with the one qubit Jaynes-Cummings model \cite{Jaynes:1963p1700}. Here the very interesting, and well studied, phenomenon of collapse and revival of qubit oscillations occurs \cite{Eberly:1980p1323}. These dynamics and their non-agreement with the semi-classical analysis present a clear indication of the very different nature of quantum and classical fields through their interaction with another quantum object. Whilst the focus of the discussion usually centres on the qubit -- the field also undergoes interesting and potentially useful dynamics. In the Jaynes-Cummings model the initial conditions are most often taken to be a coherent state of the field together with the qubit in the spin up or spin down eigenstate of the Pauli operator $\sigma_z$, or some superposition. The collapse and revival dynamics feature an interesting interplay between the field and the qubit, where these components initially entangle and \EX{\sigma_z} begins to oscillate. The initial quantum information of the qubit is then almost entirely transferred into a Schr\"{o}dinger-cat state of the field which is accompanied by  the collapse of the qubit's oscillations. In the final stages of revival this process is, to a good approximation, reversed. The information contained in the field as a macroscopically distinct superposition of states is transferred to the whole system as entanglement with the concomitant, and characteristic, revival of oscillations of \EX{\sigma_z} (see, for example, \cite{plk,Jaynes:1963p1700,PhysRevA.79.032328}). The possibilities for exploiting Jaynes-Cummings like interactions become more varied when the number of   qubits is increased to what is often referred to as the Tavis-Cummings model \cite{Tavis:1968p2457}. One particular example is that of Schr\"{o}dinger-cat state swapping between an ensemble of spins and a single field mode. Here it is possible to leverage analogies of Schr\"{o}dinger-cat states in an ensemble of spins and demonstrate that these Schr\"{o}dinger-cat states can be exchanged between the spins and the field to which they are coupled. This subject has recently been explored in depth in~\cite{Jarvis:6p2458,Jarvis:2009p2260,Rodrigues:2008p572} and given recent advances in state of the art experimental technique, as exemplified by~\cite{Baumann:2010p2463}, may soon find utility in real world quantum technologies such as for metrology. It is the phenomena of Schr\"{o}dinger-cat state swapping that we now investigate. 

There are three characteristic time scales associated with collapse and revival in the one qubit Jaynes-Cummings model. These are, for the field initially prepared in a coherent state with an average of $\bar{n}$ photons: the Rabi time given by $t_R=\pi/\left(g\sqrt{\bar{n}}\right)$ (where $g$ is the atom field coupling strength); the collapse time that sets the Gaussian decay envelope of the oscillations by $t_c= \sqrt{2}/g$; and the first revival time $t_r=2 \pi \sqrt{\bar{n}}/g$ that determines when the oscillations reappear. Importantly, at $t_r/2$, for large $\bar{n}$, the qubit and field almost completely disentangle. In the Tavis-Cummings model some of these time scales depend on the number of qubits $N$ coupled to the field mode~\cite{Jarvis:2009p2260}. Here the key observation is that the first revival time occurs at $t_{r_1}=t_r/N$. This implies that any quantum information processing operation in the Tavis-Cummings system that is based  one way or another on collapse and revival can gain a linear speedup simply by increasing the number of qubits in the system. In many experimental realisations of the Tavis-Cummings system it is the field that is most significantly effected by environmental decoherence, while the qubits can be quite long lived. The question that we address in this \PAPER is whether or not  a speedup in the Schr\"{o}dinger-cat state swap phenomena affected by an  increase in the number of qubits could be used to overcome or reduce the effects of decoherence on the field mode.

\begin{figure}[tb]
\begin{center}
\resizebox*{0.5\textwidth}{!}{\includegraphics{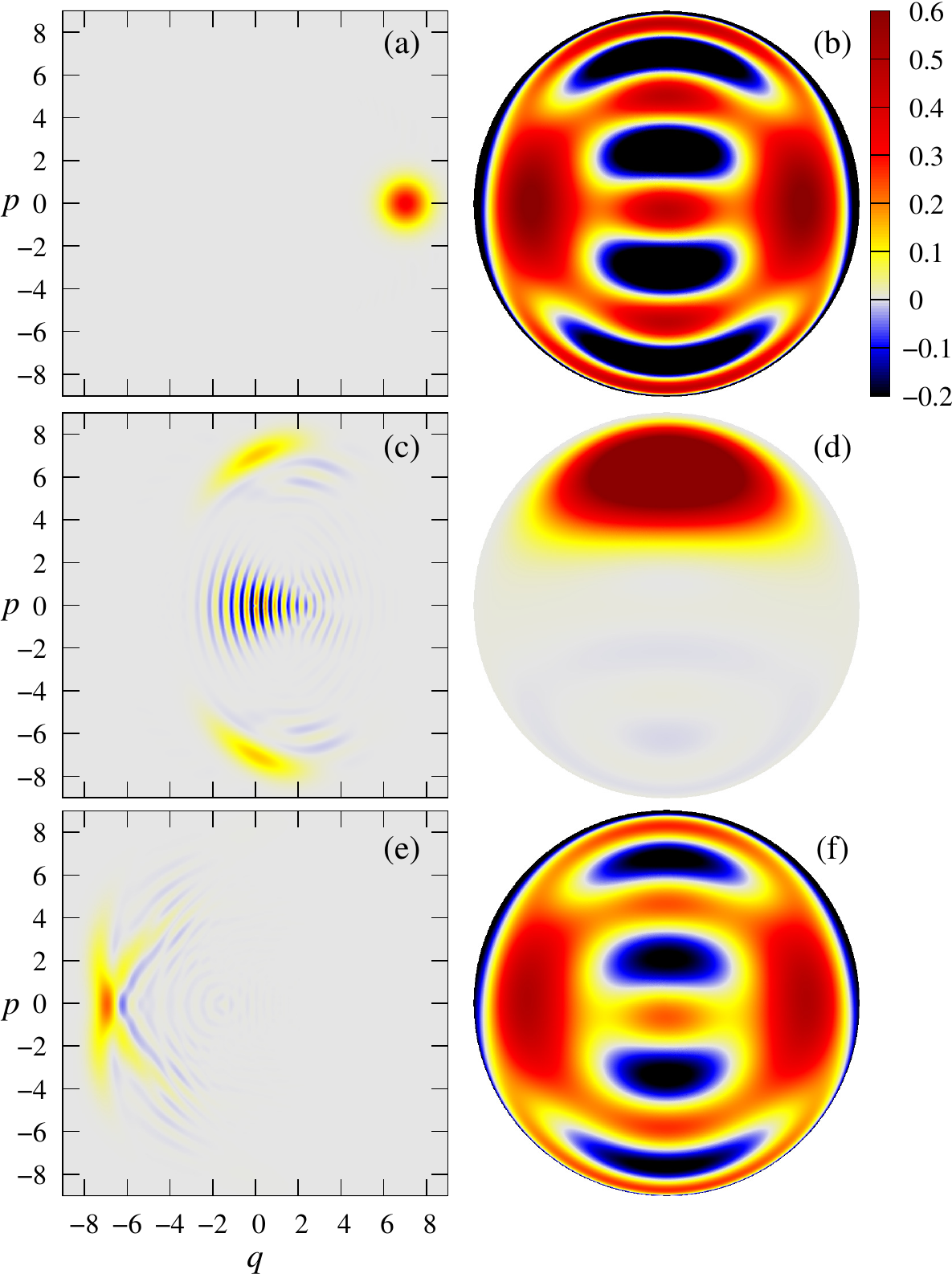}}
\end{center} \caption{(color online) The dynamical evolution of a system described by the Tavis Cummings model for an initial state of $\ket{\alpha=\sqrt{25}}\otimes\ket{\Theta(1.0,5)}$. On the left hand side we show the Wigner function of the field and on the right hand side the Lambert azimuthal equal-area projection of spin Wigner function. Snapshots are taken at (a,b) $t=0$, (c,d) $t=t_r/2N$ and (e,f)  $t=t_r/N$.}
\label{fig:catswap}
\end{figure}

The Tavis-Cummings Hamiltonian for $N$ identical qubits interacting, on resonance, via the same dipole coupling, $g$, in the rotating wave approximation with a single-mode of a quantum field can be written by extending the Jaynes-Cummings Hamiltonian~\cite{Jaynes:1963p1700,plk} and takes the following form in the interaction picutre~ \cite{Tavis:1968p2457}
\be
\label{Htc}
H =  
 \hbar g \sum_{k=1}^N  \left(\hat{\sigma}_{+}^k \hat{a}+\hat{\sigma}_{-}^k\hat{a}^{\dagger}\right).
\ee 
Here $\hat{a}^{\dagger}(\hat{a})$ is the creation (annihilation) operator for the field, with $[\hat{a},\hat{a}^{\dagger}]=1$,
$\hat{\sigma}_{\pm}^k=\frac{1}{2}\left(\hat{\sigma}_x^k \pm i \hat{\sigma}_y^k\right)$ are the qubit operators that effect
transitions between the energy ($\hat{\sigma}_z^k$) eigenstates (here $k$ is simply a qubit index), and $\hbar g$ is the coupling energy between the qubit and the field. We employ coherent states
\bel{eq:coh}
\ket{\alpha}=e^{-|\alpha|^{2}/2} \sum_{n=0}^{\infty} \frac{\alpha^n}{{n!}}(\hat{a}^\dag)^n\ket{0} \; ,
\ee
with mean photon number $|\alpha|^2$ as the initial condition for our field mode. A common way to display these states is the Wigner function  
defined in terms of position and momentum as 
\bel{Wigner1}
W(q,p)=\frac{1}{2 \pi \hbar}\int d\zeta \, \IPEX{q+\half\zeta }{\bigg.\hat{\rho}_f}{q-\half\zeta} e^{-i{p \zeta}/{\hbar} }
\ee
where  $\hat{\rho}_f$ is the density operator for the field. In \Fig{fig:catswap}(a) we plot the Wigner function for the initial state of the field \ket{\alpha=\sqrt{25}} which, as is well known, has a Gaussian profile.

An analogue of coherent states for an ensemble of $N$ spin half particles has been developed~\cite{0022-3689-4-3-009,PhysRevA.6.2211,RevModPhys.62.867}. Defining collective spin operators for the whole space of spins according to $\hat{S}_\nu=\bigoplus_{k=1}^{N}\sigma_\nu^k$ where $\nu=x, y \mathrm{\ or\ } z$, we can then define the system ground state $\ket{\mathbf{0}}$ as
the state such that $\hat{S}_z=S\ket{\mathbf{0}}$ where $S=N/2$ is the total spin of the system. This definition naturally motivates, in  analogy with the simple harmonic oscillator, the definition of raising and lowering operators for the space by $\hat{S}_\pm=\left(\hat{S}_x\pm i\hat{S}_y\right)/2$. Within this framework appears a set of states that bear a close resemblance in many of their properties to the coherent states of the harmonic oscillator \cite{Jarvis:6p2458,Jarvis:2009p2260,0022-3689-4-3-009,PhysRevA.6.2211,RevModPhys.62.867}. These take the form~\cite{Nemoto2000}
\bel{eq:sc}
\ket{z,N}=\frac{1}{(1+|z|^2)^{N/2}}\sum_{n=0}^{N}\frac{z^n}{n!}\left(\hat{S}_+\right)^n\ket{\mathbf{0}}.
\ee
These spin coherent states can be rewritten in the computational basis of eigenstates of $\{\hat{\sigma}_z^k\}$ \cite{Nemoto2000,Jarvis:2009p2260}.
\be
\ket{z,N}=\frac{1}{\left(1+|z|^2\right)^{N/2}}\bigotimes_{k=1}^{N}\left(\ket{e}_k+
z\ket{g}_k\right)
\ee
We see that a spin coherent state simply comprises a separable state with all spins pointing in the same direction. Such spin coherent states can also be represented by Wigner functions. Here we follow Agarwal et al~\cite{Agarwal:1981p2030,Dowling:1994p2031} and define the spin Wigner function on surface of a sphere by
\be
W_s(\theta,\varphi) = \sum_{l=0}^{2S} \sum_{m=-l}^l \rho_{lm}Y_l^m(\theta,\varphi)
\ee
where the total spin $S=N/2$, $Y_l^m(\theta,\varphi)$ are spherical harmonics functions and $\rho_{lm}=\mathrm{Tr}(\hat{\rho}_Q \hat{T}_l^{m\dag})$ the qubits reduced density operator. $\hat{T}_l^m$ is the multipole operator defined by:
\be
T_l^m=\sum_{n,n'=-S}^{S}(-1)^{S-n}\sqrt{2l+1} \left( \begin{array}{ccc}  S & l  & S \\-n & m & n' \end{array}\right) \OP{S:n}{S:n'}
\ee
where $\left( \begin{array}{ccc}  S & l  & S \\-n & m & n' \end{array}\right)$ is the Wigner 3j symbol and \ket{S:n} denotes the Dicke eignestate of $\hat{S}^2$ and $\hat{S}_z$. In terms of the ``spin'' basis  \ket{S:n} is simply the symmetrised sum of all states with total spin $S$ and $N_e$ spins up and $n={\left(N_e-N_g\right)}/{2}.$
For example - for three qubits and $N_e=2$ we have
$
\ket{S=\frac{3}{2}:n=\half }=\frac{1}{\sqrt{3}}\left[\ket{011}+\ket{101}+\ket{110} \right]
$.

In direct analogy with systems described in terms of the position $\hat{p}$ and momentum $\hat{q}$ operators there exist superposition states for an ensemble of spins that are macroscopically distinct, that is, Schr\"{o}dinger-cat states.  The field Schr\"{o}dinger-cat states can be represented as a superposition of two coherent states
\bel{eq:cat}
\ket{\Xi_\pm(\alpha)} = \frac{1}{\sqrt{2}}(\ket{\alpha}\pm\ket{-\alpha})
\ee
where for convenience we have assumed $\alpha \gg 1$. These Schr\"{o}dinger-cat states have spin coherent analogues of the form
\bel{eq:s.cat}
\ket{\Theta(z,N)} =  \frac{1}{\sqrt{2}}(\ket{z,N}+\ket{-z,N}).
\ee
In \Fig{fig:catswap}(b) we plot the spin Wigner function for the initial state of the spin Schr\"{o}dinger-cat state \ket{\Theta(1.0,5)}.
Here we have used the Lambert azimuthal equal-area projection~\cite{Lambert1772} of the spherical spin Wigner function where the spherical coordinates $\left(\theta_s, \varphi_s \right)$ are mapped onto the polar coordinates according to $\left(r,\theta_p\right)=\left(2\cos(\varphi_s/2), \theta_s \right)$. In this projection the north pole is the central point and its antipode is mapped onto the boundary with the equator being a concentric circle with a slightly wider radius than half that of the whole map. As with the analogous states for the Harmonic oscillator, \ket{\Theta(1.0,5)} takes the form of a a superposition of two Gaussian states (here centred on the equator) and the oscillations that are manifest between them (placed on a great circle intersecting both poles) indicate quantum coherence. It is the presence of these interference terms in Wigner functions that can be used to distinguish between macroscopically distinct superpositions of states and statistical mixtures, the latter exhibiting no such interference.

Snapshots of the evolution of the field's Wigner function and qubits' spin Wigner function under Schr\"odinger evolution for the Hamiltonian of \Eq{Htc} are shown in~\Fig{fig:catswap}. We have chosen  an initial state of $\ket{\alpha=\sqrt{25}}\otimes\ket{\Theta(1.0,5)}$  and have selected three sample times in  order to best illustrate the Schr\"{o}dinger-cat state swapping process. These are; $t=0$ (a,b), $t=t_r/2N$ (c,d) and   $t=t_r/N$ (e,f). The initial state shown in~\Fig{fig:catswap}(a,b) comprises a coherent state of the field and a Schr\"{o}dinger-cat state of spin coherent state. In~\Fig{fig:catswap}(c,d) we see that the Schr\"{o}dinger-cat state has swapped from the ensemble of spins into the field and the system's state now approximates $\ket{\Xi(\alpha\approx \sqrt{25})}\otimes\ket{z=i,5}$.  From~\Fig{fig:catswap}(e,f) at the first revival time $t=t_r/N$ we see that process has reversed  and the  cattiness of the system has swapped back from the field to the spins to form a macroscopically distinct superposition of states which is similar to $\ket{\alpha=-\sqrt{25}}\otimes\ket{\Theta(1.0,5)}$. It is this process that we now use to probe the interplay between the number of qubits in the system and environmental decoherence~\footnote{See EPAPS Document No. [number will be inserted by publisher] for an animation of the Schr\"{o}dinger-cat state swapping process using the initial condition $\ket{\alpha=\sqrt{25}}\otimes\ket{\Theta(1.0,5)}$.}. 

We will now consider the spin Wigner functions at $t=t_r/N$ as depicted, for $N=5$, in  \Fig{fig:catswap}(f). We introduce decoherence applied to the quantum field mode using a simple  Lindblad~\cite{Lindblad:1976p1778} master equation of the form
\be
\dot{\hat{\rho}}=-\frac{i}{\hbar}\left[\hat{H}_{tc},\hat{\rho}\right] + \frac{1}{2}\sum_m \left\{  \left[\hat{L}_m \hat{\rho} ,\hat{L}_{m}^{\dagger} \right] +
  \left[\hat{L}_m ,\hat{\rho} \hat{L}_{m}^{\dagger} \right] \right\}
\label{lindblad}
\ee
Schr\"{o}dinger evolution is represented by the first term and the terms due to the operators $\{L_m\}$ introduce the interaction with environmental decoherence -- such as might be introduced by coupling the system to an infinite bath of other quantum degrees of freedom. We introduce Ohmic-like damping to the quantum field via the Lindblad operator, $L=\sqrt{\Gamma} a$, where $\Gamma$ is the decay constant. This form of decoherence corresponds to the field mode being in a lossy cavity at zero temperature.

\begin{figure}[!htb]
\begin{center}
\resizebox*{0.35\textwidth}{!}{\includegraphics{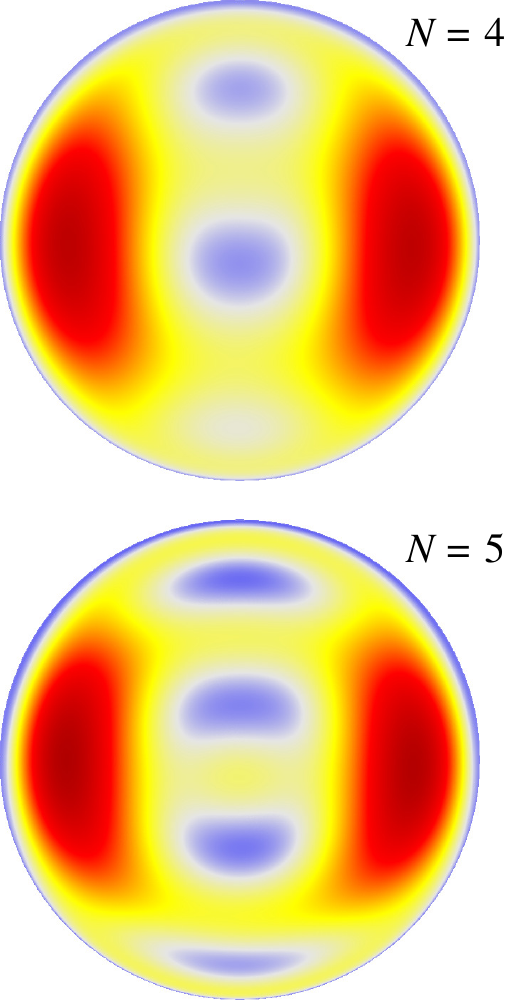}}
\end{center} \caption{(color online) Lambert azimuthal equal-area projection of spin Wigner function at $t_r/N$with $\Gamma=0.001$ for an initial state of $\ket{\alpha=\sqrt{25}}\otimes\ket{\Theta(1.0,5)}$. Data is shown for the case of four or five qubits. Here we see that the interference terms for $N=5$ are more pronounced than for $N=4$.}
\label{fig:wfd}
\end{figure}

\begin{figure}[!htb]
\begin{center}
\resizebox*{0.45\textwidth}{!}{\includegraphics{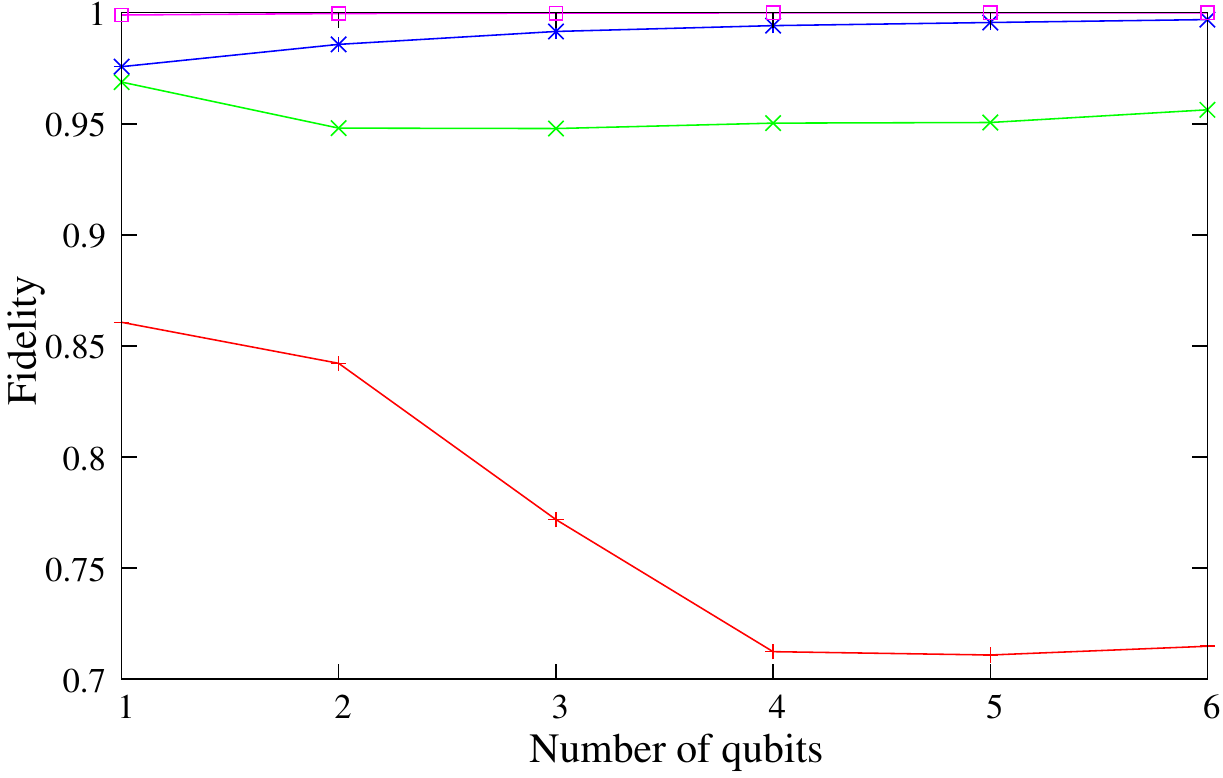}}
\end{center} \caption{(color online) Fidelity of the reduced density matrix for the spins for the decohered system as a function of the number of qubits  at $t=t_r/N$ for specific decay constants  $\Gamma =$  0.1 (red/$+$), 0.01 (green/x), 0.001 (blue/$*$) and 0.0001 (magenta/$\square$). }
\label{figfid}
\end{figure}

In \Fig{fig:wfd} we show  the spin Wigner functions at $t=t_r/N$ for $\Gamma=10^{-3}$ where we clearly see that for both $N=4$ and $5$ that the qubits are in a macroscopically distinct superpositions of spin coherent-like states. To further investigate this macroscopically distinct superpositions we plot in  \Fig{figfid} the fidelity $F=\bra{\Theta(z,N)} \rho \ket{\Theta(z,N)}$  of the resulting spin state at  $t=t_r/N$ for various $N$ and $\Gamma$. At $\Gamma=0.001$ (blue/$*$ curve) a fidelity greater than 99\% is observed with $F$ increasing as $N$ increases (at least for the limited varies of $N$ shown). Next by $\Gamma=0.1$  (red/$+$) the decoherence has become very significant with the fidelity dropping significantly as $N$ increases ($F\sim 0.72$ for $N=6$). This shows how much the quantum correlations have reduced. For smaller $\Gamma$ we observe from the strength of the interference terms that the quantum coherence between the two macroscopically distinct lumps is stronger for $N=6$ than for $N=4$.

This behavior can be explained in the following sense.  For a Schr\"{o}dinger-cat state of the form $\ket{\Xi_+(\alpha)}$ the effect of damping is to transform it to the mixed state $\rho = F |\Xi_+(\alpha e^{- \Gamma t})\rangle \langle \Xi_+(\alpha e^{- \Gamma t})|+(1-F)  |\Xi_-(\alpha e^{- \Gamma t})\rangle \langle\Xi_-(\alpha e^{- \Gamma t})|$ where   $F=(1+ \exp \left[ - 2 |\alpha|^2 \left(1-e^{- \Gamma t} \right)^2 \right])/2$. Here $t$ is the time our decoherence acts for. 
Now we know that the time for Schr\"{o}dinger-cat state swapping is determined by  the single qubit revival time and the number of qubits in the system according to $t=t_r/N=2\pi\sqrt{\bar{n}}/g N$. 
We see that as the number of qubits increases this Schr\"{o}dinger-cat state swapping time get shorter and so the field decoherence act for a shorter time. This effect wins out over the fact that the field is acting on a greater number of qubits and thus means that overall we see less dephasing induced on the Schr\"{o}dinger-cat state in the relevant time, and so a higher fidelity $F$. These field decoherence considerations have ignored independent direct dephasing acting on the qubit systems. We have done this because in effect, direct qubit dephasing is already known to be independent of $N$. If each qubit has a dephasing rate given by $\gamma$, then with $N$ qubits the total dephasing rate will be $N$ times greater. However, it is already recognised in the literature~\cite{Jarvis:6p2458} that the $1/N$ scaling in the swap time offsets this factor of $N$ in the total decoherence rate that arises from coupling $N$ qubits. This cancellation effect leaves the qubit decoherence effectively independent of $N$, so the important question is how the effect of field decoherence scales with $N$, which is what we have investigated here. Our results show that the effect of field decoherence can diminish as a function of increasing $N$, with the shortened swap time giving a genuine advantage, rather than a mere cancellation against some multiplicative factor of $N$ as for direct qubit decoherence.

To conclude, in this \PAPER we have studied in detail the transfer of quantum information between a quantum field and an ensemble of qubits within the Tavis-Cummings model. This information was encoded in terms of harmonic and spin coherent states as well as their macroscopically distinct superpositions. A spin Schr\"{o}dinger-cat state swaps to the field mode and back again at a time determined by the single qubit revival time and the number of qubits in the system according to $t=t_r/N=2\pi\sqrt{\bar{n}}/g N$. In our work we have shown that, due to this $1/N$ scaling, as the number of qubits is increased the effect of field decoherence on this swap process is reduced. We also note that in order to maintain a good Schr\"{o}dinger-cat state, where the macroscopically distinct nature of the superposition is clear, sets a practical lower bound on the field $\bar{n}$. Furthermore, experimental constraints will limit the range by which the coupling constant $g$ can be tuned. Hence, our observation that increasing the number of qubits in the system may aid in overcoming the effects of decoherence may well be of use in quantum information processing applications such as metrology.

\begin{acknowledgments}
We thank John Samson for interesting and informative discussions. 
\end{acknowledgments}

\bibliography{refTr}

\begin{thebibliography}{22}
\expandafter\ifx\csname natexlab\endcsname\relax\def\natexlab#1{#1}\fi
\expandafter\ifx\csname bibnamefont\endcsname\relax
  \def\bibnamefont#1{#1}\fi
\expandafter\ifx\csname bibfnamefont\endcsname\relax
  \def\bibfnamefont#1{#1}\fi
\expandafter\ifx\csname citenamefont\endcsname\relax
  \def\citenamefont#1{#1}\fi
\expandafter\ifx\csname url\endcsname\relax
  \def\url#1{\texttt{#1}}\fi
\expandafter\ifx\csname urlprefix\endcsname\relax\def\urlprefix{URL }\fi
\providecommand{\bibinfo}[2]{#2}
\providecommand{\eprint}[2][]{\url{#2}}

\bibitem[{\citenamefont{Wheeler and Zurek}(1983)}]{wheeler83}
\bibinfo{editor}{\bibfnamefont{J.~A.} \bibnamefont{Wheeler}} \bibnamefont{and}
  \bibinfo{editor}{\bibfnamefont{W.~H.} \bibnamefont{Zurek}}, eds.,
  \emph{\bibinfo{title}{Quantum Theory and Measurement}}
  (\bibinfo{publisher}{Princeton University Press, Princeton, NJ},
  \bibinfo{year}{1983}).

\bibitem[{\citenamefont{Penrose and Marcer}(1998)}]{Pen98}
\bibinfo{author}{\bibfnamefont{R.}~\bibnamefont{Penrose}} \bibnamefont{and}
  \bibinfo{author}{\bibfnamefont{P.}~\bibnamefont{Marcer}},
  \bibinfo{journal}{Phil. Trans. Royal. Soc.} \textbf{\bibinfo{volume}{356}},
  \bibinfo{pages}{1927} (\bibinfo{year}{1998}).

\bibitem[{\citenamefont{Nielsen and Chuang}(2000)}]{NC2000}
\bibinfo{author}{\bibfnamefont{M.~A.} \bibnamefont{Nielsen}} \bibnamefont{and}
  \bibinfo{author}{\bibfnamefont{I.~L.} \bibnamefont{Chuang}},
  \emph{\bibinfo{title}{{Quantum Computation and Quantum Information}}}
  (\bibinfo{publisher}{Cambridge University Press}, \bibinfo{year}{2000}), ISBN
  \bibinfo{isbn}{0521635039}.

\bibitem[{\citenamefont{W.J.Munro et~al.}(2002)}]{Munro2002}
\bibinfo{author}{\bibnamefont{W.J.Munro}} \bibnamefont{et~al.},
  \bibinfo{journal}{Phys. Rev. A.} \textbf{\bibinfo{volume}{66}},
  \bibinfo{pages}{023819} (\bibinfo{year}{2002}).

\bibitem[{\citenamefont{Ralph et~al.}(2003)}]{Ralph2003}
\bibinfo{author}{\bibfnamefont{T.C.}~\bibnamefont{Ralph}} \bibnamefont{et~al.},
  \bibinfo{journal}{Phys. Rev. A.} \textbf{\bibinfo{volume}{68}},
  \bibinfo{pages}{042319} (\bibinfo{year}{2003}).

\bibitem[{\citenamefont{Gerry and Knight}(2005)}]{plk}
\bibinfo{author}{\bibfnamefont{C.}~\bibnamefont{Gerry}} \bibnamefont{and}
  \bibinfo{author}{\bibfnamefont{P.~L.} \bibnamefont{Knight}},
  \emph{\bibinfo{title}{Introductory quantum optics}}
  (\bibinfo{publisher}{Cambridge University Press}, \bibinfo{year}{2005}).

\bibitem[{\citenamefont{Jaynes and Cummings}(1963)}]{Jaynes:1963p1700}
\bibinfo{author}{\bibfnamefont{E.~T.} \bibnamefont{Jaynes}} \bibnamefont{and}
  \bibinfo{author}{\bibfnamefont{F.~W.} \bibnamefont{Cummings}},
  \bibinfo{journal}{Proceedings of the IEEE} \textbf{\bibinfo{volume}{51}},
  \bibinfo{pages}{89} (\bibinfo{year}{1963}).

\bibitem[{\citenamefont{Eberly et~al.}(1980)\citenamefont{Eberly, Narozhny, and
  Sanchez-Mondragon}}]{Eberly:1980p1323}
\bibinfo{author}{\bibfnamefont{J.~H.} \bibnamefont{Eberly}},
  \bibinfo{author}{\bibfnamefont{N.~B.} \bibnamefont{Narozhny}},
  \bibnamefont{and} \bibinfo{author}{\bibfnamefont{J.~J.}
  \bibnamefont{Sanchez-Mondragon}}, \bibinfo{journal}{Phys. Rev. Lett.}
  \textbf{\bibinfo{volume}{44}}, \bibinfo{pages}{1323} (\bibinfo{year}{1980}).

\bibitem[{\citenamefont{Everitt et~al.}(2009)\citenamefont{Everitt, Munro, and
  Spiller}}]{PhysRevA.79.032328}
\bibinfo{author}{\bibfnamefont{M.~J.} \bibnamefont{Everitt}},
  \bibinfo{author}{\bibfnamefont{W.~J.} \bibnamefont{Munro}}, \bibnamefont{and}
  \bibinfo{author}{\bibfnamefont{T.~P.} \bibnamefont{Spiller}},
  \bibinfo{journal}{Phys. Rev. A} \textbf{\bibinfo{volume}{79}},
  \bibinfo{pages}{032328} (\bibinfo{year}{2009}).

\bibitem[{\citenamefont{Tavis and Cummings}(1968)}]{Tavis:1968p2457}
\bibinfo{author}{\bibfnamefont{M.}~\bibnamefont{Tavis}} \bibnamefont{and}
  \bibinfo{author}{\bibfnamefont{F.~W.} \bibnamefont{Cummings}},
  \bibinfo{journal}{Phys. Rev.} \textbf{\bibinfo{volume}{170}},
  \bibinfo{pages}{379} (\bibinfo{year}{1968}).

\bibitem[{\citenamefont{Jarvis et~al.}(6)}]{Jarvis:6p2458}
\bibinfo{author}{\bibfnamefont{C.~E.~A.} \bibnamefont{Jarvis}}
  \bibnamefont{et~al.}, \bibinfo{journal}{JOSA B, Vol. 27, Issue 6, pp.
  A164-A169} \textbf{\bibinfo{volume}{27}}, \bibinfo{pages}{A164}
  (\bibinfo{year}{6}).

\bibitem[{\citenamefont{Jarvis et~al.}(2009)}]{Jarvis:2009p2260}
\bibinfo{author}{\bibfnamefont{C.~E.~A.} \bibnamefont{Jarvis}}
  \bibnamefont{et~al.}, \bibinfo{journal}{New J Phys}
  \textbf{\bibinfo{volume}{11}}, \bibinfo{pages}{103047}
  (\bibinfo{year}{2009}).

\bibitem[{\citenamefont{Rodrigues et~al.}(2008)}]{Rodrigues:2008p572}
\bibinfo{author}{\bibfnamefont{D.~A.} \bibnamefont{Rodrigues}}
  \bibnamefont{et~al.}, \bibinfo{journal}{J Phys-Condens Mat}
  \textbf{\bibinfo{volume}{20}}, \bibinfo{pages}{075211}
  (\bibinfo{year}{2008}).

\bibitem[{\citenamefont{Baumann et~al.}(2010)}]{Baumann:2010p2463}
\bibinfo{author}{\bibfnamefont{K.}~\bibnamefont{Baumann}} \bibnamefont{et~al.},
  \bibinfo{journal}{Nature} \textbf{\bibinfo{volume}{464}},
  \bibinfo{pages}{1301} (\bibinfo{year}{2010}).

\bibitem[{\citenamefont{Radcliffe}(1971)}]{0022-3689-4-3-009}
\bibinfo{author}{\bibfnamefont{J.~M.} \bibnamefont{Radcliffe}},
  \bibinfo{journal}{Journal of Physics A: General Physics}
  \textbf{\bibinfo{volume}{4}}, \bibinfo{pages}{313} (\bibinfo{year}{1971}).

\bibitem[{\citenamefont{Arecchi et~al.}(1972)}]{PhysRevA.6.2211}
\bibinfo{author}{\bibfnamefont{F.~T.} \bibnamefont{Arecchi}}
  \bibnamefont{et~al.}, \bibinfo{journal}{Phys. Rev. A}
  \textbf{\bibinfo{volume}{6}}, \bibinfo{pages}{2211} (\bibinfo{year}{1972}).

\bibitem[{\citenamefont{Zhang et~al.}(1990)\citenamefont{Zhang, Feng, and
  Gilmore}}]{RevModPhys.62.867}
\bibinfo{author}{\bibfnamefont{W.-M.} \bibnamefont{Zhang}},
  \bibinfo{author}{\bibfnamefont{D.~H.} \bibnamefont{Feng}}, \bibnamefont{and}
  \bibinfo{author}{\bibfnamefont{R.}~\bibnamefont{Gilmore}},
  \bibinfo{journal}{Rev. Mod. Phys.} \textbf{\bibinfo{volume}{62}},
  \bibinfo{pages}{867} (\bibinfo{year}{1990}).

\bibitem[{\citenamefont{Nemoto}(2000)}]{Nemoto2000}
\bibinfo{author}{\bibfnamefont{K.}~\bibnamefont{Nemoto}}, \bibinfo{journal}{J.
  Phys.} \textbf{\bibinfo{volume}{33}}, \bibinfo{pages}{3493}
  (\bibinfo{year}{2000}).

\bibitem[{\citenamefont{Agarwal}(1981)}]{Agarwal:1981p2030}
\bibinfo{author}{\bibfnamefont{G.}~\bibnamefont{Agarwal}},
  \bibinfo{journal}{Phys Rev A} \textbf{\bibinfo{volume}{24}},
  \bibinfo{pages}{2889} (\bibinfo{year}{1981}).

\bibitem[{\citenamefont{Dowling et~al.}(1994)\citenamefont{Dowling, Agarwal,
  and Schleich}}]{Dowling:1994p2031}
\bibinfo{author}{\bibfnamefont{J.~P.} \bibnamefont{Dowling}},
  \bibinfo{author}{\bibfnamefont{G.~S.} \bibnamefont{Agarwal}},
  \bibnamefont{and} \bibinfo{author}{\bibfnamefont{W.~P.}
  \bibnamefont{Schleich}}, \bibinfo{journal}{Phys Rev A}
  \textbf{\bibinfo{volume}{49}}, \bibinfo{pages}{4101} (\bibinfo{year}{1994}).

\bibitem[{\citenamefont{Lambert}(1772)}]{Lambert1772}
\bibinfo{author}{\bibfnamefont{J.~H.} \bibnamefont{Lambert}},
  \emph{\bibinfo{title}{Beitr\"age zum Gebrauch der Mathematik und deren
  Anwendungen}} (\bibinfo{address}{Berlin}, \bibinfo{year}{1772}).

\bibitem[{\citenamefont{Lindblad}(1976)}]{Lindblad:1976p1778}
\bibinfo{author}{\bibfnamefont{G.}~\bibnamefont{Lindblad}},
  \bibinfo{journal}{Commun. Math. Phys.} \textbf{\bibinfo{volume}{48}},
  \bibinfo{pages}{119} (\bibinfo{year}{1976}).

\end{thebibliography}

\end{document}